\documentclass[sigconf,anonymous=false]{acmart}
\renewcommand\footnotetextcopyrightpermission[1]{}
\AtBeginDocument{%
  }

\setcopyright{acmlicensed}
\copyrightyear{2025}
\acmYear{2025}
\acmDOI{XXXXXXX.XXXXXXX}
\acmConference[Conference]{ International Conference for High Performance Computing, Networking, Storage, and Analysis}{Nov 15-20, 2026}{Chicago, IL}

\acmISBN{978-1-4503-XXXX-X/18/06}
\usepackage{mdframed}
\usepackage{svg}

\usepackage{xspace}
\usepackage{paralist}

\usepackage{caption}
\usepackage{subcaption}

\usepackage{mdframed}
\usepackage[T1]{fontenc} % For proper font encoding
\usepackage{minted} % For syntax highlighting
\usepackage{xcolor} % For custom colors, if desired

\usepackage{tabularx}
\usepackage{threeparttable} % For table notes
\usepackage{multirow}
\usepackage{tabularray}
\usepackage[table]{xcolor}% http://ctan.org/pkg/xcolor
\usepackage{tabularx}
\usepackage[dvipsnames]{xcolor}
\usepackage{makecell}
\usepackage[acronym,nomain,nonumberlist]{glossaries}
\usepackage{booktabs} % For professional-looking rules
\usepackage{siunitx}  % For number formatting and alignment

\usepackage[font=small,labelfont=bf,skip=2pt]{caption}
\makeglossaries

\newacronym{ai}{AI}{artificial intelligence}
\newacronym{ci}{CI}{continuous integration}
\newacronym{bsp}{BSP}{bulk synchronous parallel}
\newacronym{mpi}{MPI}{Message Passing Interface}
\newacronym{hpc}{HPC}{high performance computing}
\newacronym{aws}{AWS}{Amazon Web Services}
\newacronym{gke}{GKE}{Google Kubernetes Engine}
\newacronym{aks}{AKS}{Azure Kubernetes Service}
\newacronym{eks}{EKS}{Elastic Kubernetes Service}
\newacronym{ml}{ML}{machine learning}
\newacronym{rdma}{RDMA}{Remote Direct Memory Access}
\newacronym{os}{OS}{operating systems}
\newacronym{vm}{VM}{virtual machine}
\newacronym{llm}{LLM}{Large Language Models}
\newacronym{fom}{FOM}{figure of merit}
\newacronym{efa}{EFA}{Elastic Fabric Adapter}
\newacronym{ec2}{EC2}{Elastic Compute Cloud}
\newacronym{ucx}{UCX}{Unified Communication X}
\newacronym{cni}{CNI}{container networking interface}
\newacronym{ebpf}{eBPF}{extended Berkeley Packet Filter}
\newacronym{gvnic}{gVNIC}{Google Virtual NIC}
\newacronym{uri}{URI}{unique resource identifier}
\newacronym{api}{API}{application programming interface}
\newacronym{dag}{DAG}{directed acylcic graph}
\newacronym{crd}{CRD}{custom resource definition}
\newacronym{mcp}{MCP}{Model Context Protocol}

\newacronym{pid}{PID}{Process IDentifier}

\begin{document}

% Idea from Tapasya:
% survey of performance and usability of HPC applications in cloud
\title{Agentic Orchestration of HPC Applications in Cloud}
% \title{Cross Cloud Performance Study\protect\\ Informing Possible Futures for HPC\protect\\}

% \subtitle{\large(1) Paper Type: Regular}


\author{Vanessa Sochat}
\authornote{Corresponding Author}
\email{sochat1@llnl.gov}
\orcid{0000-0002-4387-3819}
\affiliation{%
  \institution{Lawrence Livermore National Laboratory}
  \city{Livermore}
  \state{California}
  \country{USA}
}

\author{Daniel Milroy}
\email{milroy1@llnl.gov}
\orcid{0000-0001-6500-3227}
\affiliation{%
  \institution{Lawrence Livermore National Laboratory}
  \city{Livermore}
  \state{California}
  \country{USA}
}

\renewcommand{\shortauthors}{Sochat and Milroy.}

\begin{abstract}
Large Language Models (LLMs) are serving as a catalyst of change for research practices, touching the daily lives of staff scientists, software engineers, and system administrators. The developments promise new degrees of autonomy, where categories of human work and decision making are replaced by autonomous, goal-oriented systems. This transition necessitates novel architectural paradigms and solid understanding of the strengths and limitations of LLMs.  In this work, we design agents to intelligently deliver the entire life-cycle of an HPC application experimental run in cloud -- creation and build of a container, deployment in Kubernetes, optimization, and orchestration of a scaling study. We pursue this task for four well-known HPC applications to build multi-platform images and optimize across 21 instance types in Kubernetes. We demonstrate successful linear scaling with patterns approved by human experts, designs that improve work time to completion, and review suggested best practices for agentic design and collaboration.

\end{abstract}

\maketitle
\section{Introduction}

The revolution of \gls{llm} for usage in \gls{ai} and \gls{ml} workloads has taken the global research community by storm. The \gls{hpc} community is a representative subset of this user base that can benefit from using \gls{ai}/\gls{ml} models to advance science. The interfaces to interact with \gls{llm}s are typically agents -- applications that can receive instructions and respond with context-aware, meaningful text in coordination with \gls{ai} services \cite{llm-agent-definition}.  Successful integration of agents into scientific workflows depends on understanding strengths and limitations, and strategy to holistically combine human goals with \gls{llm} productivity. A \gls{llm} is only as powerful as its ability to focus on a scoped task, and an agentic team of \gls{llm}s and humans will best achieve a desired outcome with proper protocol for guidance and validation. As natural language becomes a part of a new type of software to create revolutionary new systems, the feedback loop between agents and humans must become tighter \cite{Combinator2025-gq}.

A variety of developer frameworks \cite{google-agent-development-kit,azure-ai-foundry} and hosted services are available for inference, including Google's Gemini, Anthropic's Claude, OpenAI's ChatGPT, and Microsoft Copilot. These \gls{ai} systems are being used in medicine, education, and science \cite{wei2024evaluation,laun2025chatbots,Jr2023-jr}. They are designed to derive objectives from prompts, and synthesize multiple data sources while using internal tools to plan, reason, and execute complex, multi-step tasks. While the structure is not typically transparent to the user, we can speculate that these systems have checks and balances, can adapt to failure, and use a mixture of experts (MOE)~\cite{Gan2025-hx} to target sub-tasks to models with tuned expertise. Although a response to the user may appear as a single, cohesive output, it likely results from synthesis of multiple sources of information and multiple steps, each with support from tools. Tools might include symbolic solvers for formal logic, code interpreters to produce accurate mathematical results, and search engines or databases for general knowledge retrieval. A planner that initially processes the prompt is likely responsible for orchestrating agents and tools into an execution pipeline. Industry vendors are ahead of the game to structure these interactions, developing the \gls{mcp} to standardize \gls{ai} system interactions \cite{openai-api}.

While corporations can afford and utilize these services, scientific groups are more cost constrained and can be limited by institutional regulations. While downloading models and using them on-premises might be allowed, the additional tools and agentic \gls{ai} systems are not included. The inability to fully utilize the latest \gls{ai} systems hinders the ability of the scientific community to use them for advances and learning. % Here I mean we would be slower in being able to realize the gains from the approach. A researcher with access today will mode faster toward solving a problem than a researcher than has to first wait for an institution to deploy something much lesser.
The requirement to deploy models locally is redundant, and adds management complexity along with capital and operational costs. However, the challenges to using \gls{ai} for computational science does not deem the task impossible. Scientists can develop infrastructure and software approved for scientific institutions. These needs are demanded by national initiatives such as FG-HPCC and Genesis \cite{Gil2025Genesis,llnl_fghpcc_rfi_2025}. A shift from laborious manual task execution to proactive, goal-oriented autonomy would be a considerable benefit to scientific discovery.

\subsection{Agentic \gls{ai} for Computational Science}

% I'm not sure I have evidence to back up this claim... it just seems obvious to me that it does/will even more.
Research specific to agents oriented to build, deploy, and optimize \gls{hpc} applications is in its infancy. We might imagine a future when \gls{ai} systems are used as first class citizens for scientific workflow development and execution. 
%It is unclear how agentic steps could integrate with traditional \gls{hpc} workload managers that rely on static, heuristic-based approaches or graph-based algorithms, often requiring substantial time to wait in a queue that would further exacerbate the ability of multiple agents to work together in real-time. However, 
In that many agentic \gls{ai} systems decompose high-level tasks into directed graphs of general computational resources, systems that support agentic orchestration \cite{UnknownUnknown-me} or graph-based schedulers \cite{AHN2020202} could match resources to agent needs. % offer promise as frameworks to match resources to agent needs. % I'm not sure this is related but I had to throw it in :) We should either remove it or find a better way to mention it, or save it for a future paper.
Work to integrate \gls{ai} into scheduling is already years old, however it is often focused on prediction of a specific attribute such as runtime remaining for the job \cite{Wang2021-mr}. An agentic framework to orchestrate a multi-step workflow that might require coordination with humans adds additional complexity to the task. Agents may not yet be able to act independently as an~\gls{hpc} scheduler, but they can optimize resource requests for a specific application. 

% This is the line I was trying to get to - 

Scientific workflows often require manual tasks to prepare for using workflow tools. For example, running a Snakemake \cite{Molder2021-rq} pipeline might require building a container first. Directed acyclic graph-oriented workflows are typically deterministic without \gls{llm} routing \cite{Pranoy2025-ba}. An agent-based approach could help to make this process flexible and automated, assisting with preparatory steps, runtime parameters, resource requests, topology, and binding.

Agentic \gls{ai} offers the opportunity to re-imagine this sequence of steps. While many scientific-based \gls{ai} tools focus on hypothesis generation, data analysis, experiment design, and scientific writing \cite{gridach2025agentic,Datar2025-ud}, a less studied and more challenging area of work is in the space of workload orchestration. A successful agentic framework for executing workload steps would require collaboration between agents and human experts to reach a desired outcome. Akin to hosted services \cite{Salomone2025-dl} and workflow tools, a top level manager would handle decision making, supplemented by human intervention to further validate experiment progress. In our work, we prototype and study such a tool -- an agentic team of step-level experts orchestrated by a top level manager and supporting sub-agents to achieve a full build, deploy, and optimization of \gls{hpc} applications. We make the following contributions:

\begin{compactitem}
 \item{Software prototype for agentic orchestration}
 \item{Agentic autoscaling for instance selection across 21 types}
 \item{Execution and analysis of 5 HPC applications in Kubernetes}
 \item{Comparison of outcomes against human expertise}
 \item{Best practices for \gls{ai} and human collaboration}
\end{compactitem}

We start with an introduction to agentic roles and definitions (Section \ref{sec:methods}) and describe our methods to orchestrate agentic execution and respond to failure. We describe experiments and results (Section \ref{sec:results}). We finish with a discussion of lessons learned and best practices for collaboration between agents and humans.
\section{Methods}
\label{sec:methods}

\label{sec:overview}
\subsection{Overview}
We aimed to orchestrate the entirety of a build, deployment, and optimization cycle for 5 \gls{hpc} applications of interest in the leading cloud orchestration framework Kubernetes \cite{kubernetes}. We chose Kubernetes because it provides declarative management and structured, programmatic interactions that can be easily used by agents. Our applications include a set that vary in difficult to build, including LAMMPS, AMG, Kripke, the OSU Benchmarks, and Laghos, all of which we have deployed and described in previous work \cite{sochat2025usabilityevaluationcloudhpc}. For our setup, to give the optimization step a choice of resources, we aimed to use an autoscaling setup on \gls{aws} that includes 21 instance types (Table \ref{table:instances}) that include each of arm64 and amd64 platforms. Instance sizes were chosen to be approximately \$3.00 or less. Providing a dimension across platforms, micro-architectures, CPU, and memory would give the agents four dimensions to consider when selecting an instance type.

% 91 ARM (no gpu)
% 126 CPU (no gpu) (total for CPU is 217)
% 119 GPU for equivalent x86.
% total is 336

\begin{table}[htbp] % Added placement specifiers for better float handling
 \centering
 \begin{threeparttable}
 \small
  \caption{Instance Types for Agentic Selection}
  \label{table:instances}
  \begin{tabular}{lllll}
      \toprule
    Instance & Processor & Cores/Freq. & Mem. & Cost/Hr \\
    \midrule
     c6a.16xlarge & AMD EPYC 7R13 & 32/3.6GHz & 128GB & \$2.448 \\
     c6i.16xlarge & Intel Ice Lake & 32/3.5GHz & 128GB & \$2.72\\
     c6id.12xlarge & Intel Ice Lake & 24/3.5GHz & 96GB &  \$2.4192 \\
     c6in.12xlarge & Intel Ice Lake & 24/3.5GHz & 96MB & \$2.7216 \\
     c7a.12xlarge & AMD EPYC 9R14 & 24/3.7GHz & 96GB & \$2.4634 \\
     c7g.16xlarge & AWS Graviton3 & 64/2.5GHz & 128GB & \$2.32 \\
     d3.4xlarge & Intel Cascade Lake & 8/3.1GHz & 128GB & \$1.998\\
     hpc6a.48xlarge & AMD EPYC 7R13 & 96/3.6GHz & 384GB & \$2.88 \\
     hpc7g.16xlarge & AWS Graviton3 & 64/2.6GHz & 128GB & \$1.6832 \\
     i4i.8xlarge & Intel Ice Lake & 16/3.5GHz & 256GB & \$2.746 \\
     m6a.12xlarge & AMD EPYC 7R13 & 24/3.6GHz & 192GB & \$2.0736 \\
     m6g.12xlarge & AWS Graviton2 & 48/2.5GHz & 192GB & \$1.848 \\
     m6i.12xlarge & Intel Ice Lake & 24/3.5 GHz & 192GB & \$2.304\\
     m6id.12xlarge & Intel Ice Lake & 24/3.5GHz & 192GB & \$2.8476\\
     m7g.16xlarge & AWS Graviton3 & 64/2.5GHz & 256GB & \$2.6112 \\
     r6a.12xlarge & AMD EPYC 7R13 & 24/3.6GHz & 384GB & \$2.7216 \\
     r6i.8xlarge & Intel Ice Lake & 16/3.5GHz & 256GB & \$2.016 \\
     r7iz.8xlarge & Intel Sapphire Rapids& 16/3.9GHz & 256GB & \$2.976 \\
     t3.2xlarge & Intel Skylake & 4/3.1GHz & 32GB & \$0.3328 \\
     t3a.2xlarge & AMD EPYC 7571 & 4/2.5GHz & 32GB & \$0.3008\\
     t4g.2xlarge & AWS Graviton2 & 8/2.5GHz & 32GB& \$0.2688 \\     
  \bottomrule
\end{tabular}

\begin{tablenotes}[flushleft]
    \footnotesize % Use a smaller font for notes
    \item Cloud instance types available for selection.
    \item Instances were selected to be under \$3.00 to provide a 4 dimensional gradient.
\end{tablenotes}

 \end{threeparttable}
\end{table}

\label{sec:agents}
\subsection{Agents}

We designed an agentic framework with a preference for simplicity. While several libraries exist to use \gls{mcp} \cite{model-context-protocol}, for our own learning we chose to minimize external dependencies and maximize transparency by designing each agent with a common base class to query the Gemini \gls{api} and take on a specific role. While most \gls{mcp} agent \gls{api}s are expecting request and responses exclusively in text, we wanted our design to be oriented around a directed graph, and based on execution of controlled commands that resulted in a clear result (e.g., a return code) that would not require returning to an \gls{llm} to complete an interaction. In this controlled environment, multiple agents that work together form an agentic system. Agentic systems contain different types of agents that operate at different path lengths from the root. Together, they form an agentic graph.

\smallskip
\noindent{\bf Step Definition} 
\label{sec:steps}
During the execution of a workflow, multiple tasks are connected by inputs and outputs and end in a final state. Each task accepts inputs, and executes a function on the inputs to transform them into outputs. Inputs, outputs, and processing together define a step. The inputs and outputs between two adjacent steps form a shared context. Multiple steps (nodes) that have dependency relationships (edges) and a shared context form a graph. In this graph, a step A is adjacent and directed from A to B if the output of step A serves as the input to step B.  As an example, the resulting container \gls{uri} is an output of a build step, and could be used as input to a deployment step. 

\smallskip
\noindent{\bf Step Agent} 
\label{sec:step-agent}
A step agent is responsible for a specific task. A single step agent can act as an independent unit or entity, and can be represented as a node in a graph. Two step agents can be adjacent in the graph and joined by an edge if their inputs or outputs are compatible. For example, a build agent builds an image before it is provided to a deploy agent to test. The structure of any step agent is simple: a controlled set of tasks specific to the agent that are combined with structured prompts. The prompts are scoped to derive input (parameters, configuration files, or supporting regular expressions) to direct a task. Input comes by way of the context (Section \ref{sec:context}), which includes pre-defined variables specific to the task at hand (e.g, a container \gls{uri} for a build agent) and a general \emph{details} section where a user can provide freeform text instruction to the agent. Each step agent is implemented with actions and validation checks for the task. For example, a build agent takes a response from the \gls{llm} and writes a Dockerfile to a temporary staging directory, and collects output and error to return as feedback to the \gls{llm}. A return code of ``0" indicates success and proceeds to the next step. A deploy agent is then provided with the Dockerfile and tasked to generate a YAML manifest to deploy it. Step agents use a conversational client, so new attempts are made with memory (model context) of previous failures. Each agent instantiates its own \gls{llm} model, and this is done so that responsibility does not bleed between agents. For our agentic system, we define step agents for a Dockerfile \emph{build}, a Kubernetes deployment for a \emph{Job} and Flux Framework \emph{MiniCluster}, an \emph{optimization} task, and a \emph{scaling} decision. 

% The indentation looks the same as the above (and its a subsection) but that is probably ok.

\paragraph{Build Agent} The \emph{build} agent is an expert at building Dockerfiles scoped for an \gls{hpc} application, and is allowed to design the Dockerfile to be given to a \emph{docker build} or \emph{docker buildx} command in a sandbox on the user system.  The execution of the build is done by the agent's class, and provides the logic to check the return code, and either return to the manager to proceed to the next step, or in the case of a build error, send the output to a debug helper agent to diagnose the issue. The debug helper agent then provides scoped feedback to the build agent to retry. Multiple build attempts form a cycle in the graph, and a successful build determines breaking the cycle and moving on to the next step. Context variables for the build agent include a \emph{container} \gls{uri}, an \emph{application} to build,  an \emph{environment} and \emph{platforms} to build for, and whether to \emph{push} or \emph{load} the image into Docker. A push requires the build environment to have permissions for the task, and the build agent is only allowed retries up to a maximum number of attempts for each step.

\paragraph{Deploy Agent} The \emph{deploy} agent is responsible for the single task of taking an instruction for an application of interest and a Kubernetes abstraction (e.g., Job or Flux Framework MiniCluster \cite{Sochat2024-the-flux-operator}) and successfully running the application in Kubernetes. The agent is instructed to deploy a container for a specific environment and application, and with additional context provided in the details section. The deploy agent class validates the manifest generated by attempting to load it into YAML, and then checking that the container \gls{uri} matches what was requested. The class then writes the manifest to a temporary deploy directory, and applies it to the cluster with \emph{kubectl}. The details in the manifest including the container, command, size, and resources, are decided by the \gls{llm} and can be influenced by details in the prompt. The application Pods are  monitored, and the class looks for changes of state from \emph{Running} to \emph{Completed} or \emph{Failed}. A maximum waiting time for a monitored Pod can be set by the user, which allows for handling cases such as timeout due to error or insufficient resources. The class can handle edge cases such as timeout, out of memory errors \emph{OOMKIlled}, or in the case of Flux, the request being unsatisfiable. The class also handles unexpected deletion. In the cases of an error code, the error along with cluster diagnostics are sent to a debug agent to determine and summarize the issue to inform the deploy agent how to retry. The deploy is attempted up to a maximum number of attempts or a successful case (\emph{Succeeded} status).

\paragraph{Optimization Agent} The optimization agent can be viewed as a child of the deploy agent in that it is run after a successful deploy to improve upon a metric of interest. The metric of interest must be defined in the context details -- an \emph{optimize} variable -- that triggers its creation. The optimization agent works as follows. A completed result log and manifest are passed forward in the context from a deploy step. The prompt instructs the agent to optimize for the metric of interest, and return a JSON structure with variables to update the previous manifest. For the first execution, there is no means for the optimization agent to know what the metric of interest value is from the provided log, and so a child helper agent class is instructed to generate a regular expression to parse the metric of interest from the log. The result parsing helper agent receives a regular expression back from the \gls{llm}, and tests it against the log. Validation consists of ensuring there is a match, and then asking for human validation. The user deploying the pipeline is given the task to say it is correct (yes), incorrect (no), or to provide custom feedback (feedback). The result parser agent runs in cycles until a successful result -- one that passes automated checks and is approved by the human assistant.  The result is cached and reused. 

The optimization agent \gls{llm} is provided with the current and all previously parsed metrics of interest, the latest log and manifest, and instruction for how to optimize from the user. The agent is instructed to return a newly optimized manifest in JSON. Each subsequent retry assesses performance of deploys before it. A \emph{kubectl explain} is done for the resource of interest (e.g., Job or MiniCluster) to provide the exact fields and descriptions of the manifest that can be changed, and the Kubernetes Python SDK is used to get a summary of nodes (counts and resources) that are available for using. A \emph{resources} field in the plan is also allowed for the user to provide explicit instructions about what resources to use. Such a field is useful for autoscaling clusters, where nodes available might not be present in the cluster. The optimization agent is required to return an updated manifest, a decision to \emph{RETRY} or \emph{STOP}, and a reason. 
%The optimization agent is triggered to run when the user has defined optimization instructions in the plan for the step. The design of the optimization agent is clever in that to execute a test of a new configuration, it returns a call to the deploy agent, however, editing the context to indicate an optimization is in progress. An optimization in progress changes the execution flow to return to the optimization agent on a successful deploy. 
In the case of error, execution returns to the debug agent and then deploy to resolve it before optimization is continued. While the current implementation uses the optimization for Kubernetes, it is not tied to Kubernetes. The same optimization agent could be used by a bare-metal execution deployment agent and tasked equivalently.

Due to common interactions with Kubernetes to get logs or status, the two deployment agents share a common set of functions and underlying Python class. These controlled interactions are how agents access the execution environment. The orchestration software  carefully monitors each execution and determines when the state is erroneous. For example, deploying an abstraction to Kubernetes could fail immediately if the YAML file is invalid, or later if execution starts and fails. Collection and filter of error information from the right sources is a strategic task that ensures that a failed attempt can be debugged properly. Direction of the next agent to make a request to with a specific update to the prompt is essential. For example, not capturing a timeout or out of memory issue can lead to updates to the manifest that are nonsensical like increasing the problem size. In the case of failure, a cleanup is typically required to prepare for the next attempt.

\vspace{5pt}
\noindent{\bf Prompt Structure} 
\label{sec:step-agent}
Each agent has scoped and structured prompts to perform and retry a task. A prompt defines a clear persona, context, task, and set of instructions. The persona clarifies the agent's role. The task and context clearly define the goal for the interaction. The instructions are concise, single lines that describe what an agent \emph{MUST} and \emph{MUST NOT} do. Using capital letters to provide emphasis makes a difference. Not adding this emphasis leads to agents more often not following instructions. Using regular expressions to filter text noise from prompts improves goal attainment. For example, when debugging output from the build agent, many thousands of lines are generated by apt-get during an Ubuntu build. These extra lines distract from identifying the core issue and can be filtered before sending to Gemini. This reduced our token count (Section \ref{ref:token-counts}) from the order of 300,000 tokens down to a few thousand, and reduced time to parse the response and cost.

\smallskip
\noindent{\bf Helper Agents} 
\label{sec:helper-agent}
A helper agent is an agent used by a step that is primed to perform a specific task, and typically one that does not require conversational memory. When agents are created, \gls{api} clients are instantiated with the class that can either send one-off model messages, or a message to continue a conversation. We consider the conversation a form of superficial memory, as subsequent prompts are informed by previous ones. In the case of helper agents, we would not want any bias based on a previous prompt. Our two use cases for helper agents were debugging and synthesizing errors (i.e., identifying the issue and summarizing it succinctly), and parsing a result from a log. We call these helpers the \emph{debugging} and \emph{result} agents, respectively. % In addition to summarizing and suggesting a fix for an error, the debugging agent can return execution flow to the manager if it determines that the step in question cannot resolve the issue. For example, a missing library in a container cannot be resolved by a deploy agent; the build agent must fix the underlying container build. The result agent is used by the optimization step agent to parse a metric of interest from a log. The result agent is prompted to generate a regular expression that is applied to the log. When a match is returned, it is interactively presented to the running user to validate correctness. A working regular expression only needs to be derived once for a run, as it can be validated and used for subsequent parses. This is an example of a human-validated step, and collaboration between the running human user and an agent. We found this kind of human input was needed to validate the output of the agentic task. 
The pattern of an agent performing a task and asking for human validation is one that can be extended to other use cases.

\begin{figure}[t!]
    \centering
    \includesvg[width=0.8\linewidth]{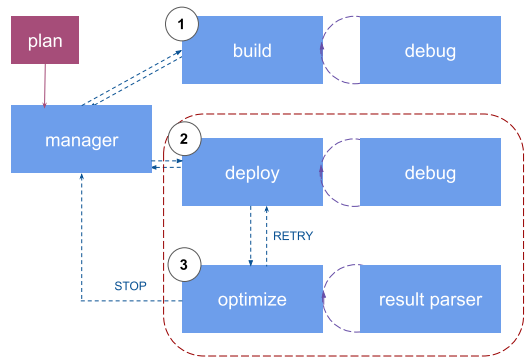}
    \vspace{1em}
  \caption{Agentic \gls{ai} System. \normalfont The manager receives a plan from the user, a YAML specification with a series of steps and context for each. The manager orchestrates executing steps defined in the  plan, allowing step agents to work independently up to a maximum number of attempts, and return to the manager if decided by a helper agent or on success. The optimization works with the deploy agent until a decision is made that the run is optimal.}
  \label{fig:agentic-ai-team}
\end{figure}

\smallskip
\noindent{\bf Manager} 
\label{sec:manager-agent}
An agentic team requires a manager to orchestrate the sequence of steps. The manager takes as input a plan derived by a human defining a sequence of steps. The manager is responsible for orchestration of step agents. Each step has a defined context. For example, the build agent minimally requires an application name and environment to tune the build for. The manager is the root of the graph, executing individual steps, and having flow return to it in the case of error or completion.  Each step has a maximum number of attempts, or up to the point when a debugging agent directs flow back to the manager. The manager saves an output file with metadata when the workflow completes. The manager combined with step agents and helper agents form an agentic team (Figure \ref{fig:agentic-ai-team}).

\smallskip
\noindent{\bf Context} 
\label{sec:context}
A shared context object is responsible for sharing state between agentic steps, and provides an interface for setting and retrieving input variables. Each step requires a specific set of inputs to be provided in the context, and inputs can be populated in several ways. The developer user adds a function to the agent class that returns a subparser. Any argument defined there is added to the agent context, and can be customized via the command line, the plan YAML file, or by an agent setting it directly on the context. As an example, a build agent by default will generate and set a \emph{container} \gls{uri}. When the context is passed to the deploy agent, the input \emph{container} is not only instructed for use, but validated by the agent class to appear in the output. The validation ensures that cycles are not wasted testing a \gls{uri} that does not exist. A user might also create a plan for a previously built container, and provide the \gls{uri} as the \emph{container} field under the context. This design allows for each agent to define a different set of context variables as inputs, and flexibility for them to be set by humans or another agent.  The manager loads the initial context from the user-defined plan, and each step is provided with the context for execution. Common variables between steps that aid with communication include a final \emph{result} for the current step, and an \emph{error\_message}. % For example, when a build is erroneous, the full output is given to the debugging agent with instructions to process and summarize the issue. The more succinct description and suggested fix is then passed back to the build agent for another attempt. The build agent uses a conversational client so context about previous attempts is carried forward for subsequent attempts.

\smallskip
\noindent{\bf Explainability} 
\label{sec:explainability}
The extent to which we can track and explain the decisions of the \gls{llm}s is the degree of explainability \cite{prov-agent}. We found it useful to display inputs, outputs, and prompts in the terminal, and always prompt the \gls{llm} agents to provide rationale for choices. While this is not standardized collection of provenance \cite{prov-agent} it allows us to track what is going on at a level appropriate for being able to improve upon it. Visual monitoring is useful to get direct feedback about what prompt or input data led to a decision, what information might be missing, and what factors led to surprising or improper behavior. In addition to terminal output, the manager saves all intermediate steps and output generated. Each step agent saves custom metadata, function timings, and relevant metrics of interest.

\smallskip
\noindent{\bf Controlled Interfaces} 
\label{sec:manager-agent}
We use the Google Gemini \gls{api} as our \gls{llm} backend. Each agent has defined environment interfaces that use the Python \emph{subprocess} library to execute commands for building or deploying configuration files. We typically do not allow agents to write commands that are executed to the system, but instead to populate the content of configuration files used. We currently do not provision connectors to other external databases, search engines, or \gls{api}s to supplement the agent. The closest that the \gls{llm} will get to code execution is writing regular expressions for log parsing to be used by the result agent. We looked into a potential security risk with allowing execution of regular expressions, and saw the risk of a ReDOS \cite{redos}. From a hosted service like Gemini provided by Google we saw this as unlikely, and if it happened, would be a bug in a product to report to Google. As an extra layer of precaution, all experiments are run on cloud virtual machines. These controlled interfaces are essential for agents to interact with and adapt to the environment.

% 10/31/2025 note to Dan - not sure if the last sentence above is too human-y, but I liked the idea of the agent learning from the environment so I kept it. If "learn" is too strong, we can just say "interact wtih the environment." (feel free to change)

\label{sec:experiments}
\subsection{Experiments}

\smallskip
\noindent{\bf Single Node Experiments} 
\label{sec:single-node-experiments}
We aim to build, deploy, and optimize 4 \gls{hpc} applications in Kubernetes on a single node. We provide instructions in the plan to use the Flux Operator \emph{MiniCluster} \gls{crd} \cite{Sochat2024-the-flux-operator} to deploy an \gls{hpc} cluster to run each application. We give the optimization agent step 21 types of instances (Table \ref{table:instances}) to select from, and request deployment on one node to maximize \gls{fom}. We start with one node to assess performance of the agentic team when scale is not a factor. For each of LAMMPS, AMG, Laghos, and Kripke, we give the optimization agent instructions for how to make a decision to retry or stop, along with an instruction for an optimization strategy. For all applications we provide instructions to use a smaller problem size for testing, and increase to a larger size for the optimization. In all cases, the optimization agent step is required to return a decision value of \emph{RETRY}, or \emph{STOP} with a reason and resources if a retry is needed. We test the following optimization strategies, each of which instructs the agent to maximize the \gls{fom} with additional instruction.

\begin{itemize}
\item \textbf{\gls{llm} decision}: Optimize \gls{fom} and decide when to retry/stop
\item \textbf{User function}: Require the agent to execute a function.
\item \textbf{User guided function}: Require calling a function with scaling metrics to get back optimization hints.
\end{itemize}

For all optimization strategies, we provide the agent with a description of resources available, including instance types (CPU and memory) in the autoscaling cluster. For the first optimization strategy, \emph{llm decision}, we leave the entire decision up to the agentic model. This strategy can be considered free-form in that the prompt does not give the \gls{llm} an instruction or algorithm for making a decision. For the second strategy, \emph{user function}, we provide a function in the prompt that calculates a configuration to use for the next step, and ask the agent to run it. This strategy is conceptually similar to using an \gls{mcp} server in that it mimics the \gls{llm} executing a function and using the exact output. The third strategy is an intermediate between these two extremes, taking in the current resources and returning a decision to \emph{RETRY} or \emph{STOP} with a strategy hint, and providing instruction to also return resources and instance type.  The main difference between a user-function and a user \emph{guided} function is that the latter returns metrics that the \gls{llm} can use to make its own decision. % Specifically, the agent runs a function to derive the performance ratio, scaling efficiency, memory utilization, and application-specific metrics to help guide the decision.

\smallskip
\noindent{\bf Multi-Node Experiments} 
\label{sec:multi-node-experiments}
We perform a modified version of the single node experiments that instruct the agent to do a build and deploy across multiple (N=4) nodes. We decided to use the \emph{m7g.16xlarge} due to reasonable cost and best availability across the top 3 contenders as determined by testing. Importantly, the selected instance needed to support the \gls*{aws} \gls*{efa} suggested for \gls*{hpc} workloads \cite{amazon-efa}. \gls{efa} allows network packets to bypass the operating system and go directly to the device for lower latency. The agents are instructed to build OpenMPI with libfabric intended for \gls{aws} \gls{efa}, and that they should maximize \gls{fom}. We will use the best performing strategy identified in Section \ref{sec:single-node-experiments}. Since we are running across nodes, we will add in the OSU Benchmarks in place of Laghos to test the ability of the \gls{llm} to figure out point to point and collective \gls{mpi} calls. 

\smallskip
\noindent{\bf Scaling Study} 
\label{sec:scaling-study}
For a final experiment, we challenge the agents to complete the entirety of a scaling study. For this work, we test LAMMPS, Kripke, and AMG, and pin the container build to a known working variant from our previous experiments. A scaling agent is added to the agentic team, and is prompted to return a response that decides when to continue or stop scaling toward a user-defined goal (Figure \ref{fig:scaling-study}). The optimization agent is provided a modified prompt with the context of the scaling study that instructs to change the problem size only at the first size, and then hold it constant (strong scaling). % or adjust the global problem size to have a constant amount of work per processor (weak scaling). 

An example optimize and scale plan would define a YAML file to run scaling and optimize agents run under the deploy (MiniCluster) agent. The execution of the deploy agent is allowed a maximum number of 10 attempts, and an execution timeout of 300 seconds. The container provided was previously built by a build agent, and provided to save time and not rebuild each time. The environment definition is minimal, as the details for the cluster size and resources are provided programmatically. To ensure that the execution loop has fewer edges, no returns are allowed to a human or to the manager. The scale agent is instructed to maximize \gls{fom} and pin a problem size from the smallest size (N=1) and to stop when the application stops strong scaling. The optimize instruction is specific, placing emphasis on points with \emph{MUST}. % Emphasized instructions typically result from an agent improperly performing on a testing attempt. 

% \begin{mdframed}[linecolor=white, topline=true, bottomline=true, leftline=false, rightline=false, backgroundcolor=white]
%\begin{minted}[
%    framesep=2mm, % Separation between frame and code
%    bgcolor=White,
%    fontsize=\footnotesize, % Font size of the code
%    linenos, % Show line numbers
%    breaklines % Allow lines to break if too long
%]{yaml}
%name: Scaling Study for LAMMPS
%description: Scale lammps up to 5 nodes.
%plan:
%- agent: minicluster
%  context:
%    environment: "AWS CPU instance in Kubernetes" 
%    container: ghcr.io/converged-computing/fractale-agent-experiments:lammps-reax
%    max_attempts: 10
%    max_runtime: 300
%    allow_return_to_human: false
%    allow_return_to_manager: false
%    sizes: [1,2,3,4,5]
%    scale: |
%      Strong scale lammps to maxime the FOM and minimize running time at each size.
%      You MUST choose a problem size at the smallest size (1) that you keep constant.
%      Stop when you determine the application is no longer strong scaling.
%    optimize: |
%      You MUST maximize the LAMMPS FOM, *atom steps per second.  
%      When parsing the log you MUST consider it could be Matom or katom.
%      You MUST NOT change parameters after the first scaling size (1).
%      You MUST increase problem size at the first scaling size (1) until the %job times out.
 %     You MUST increase the problem size ONLY at the first scaling size.
 %     At large sizes, you MUST adhere to the same resource limits, requests, and selectors.
%      You MUST adhere to ONE node (64 tasks) for the initial optimization.
%\end{minted}
%\end{mdframed}

The study is a collaboration between the scaling, deploy, and optimization agents, where the optimization agent derives the configuration, the deploy agent is tasked with running work at each subsequent size, and the scaling agent response determines when to stop. A human is asked to intervene to validate the parsing of a \gls{fom} from the log. We plan to instruct the agent to build and deploy a multi-node cluster starting at size two up to a maximize size of 32 nodes, and to choose the instance type for each that performed optimally for the multi-node experiments. We instruct the agents to stop when strong or weak scaling has ended, and instruct that each decision be explained with evidence from relevant literature.

\begin{figure}[b!]
    \includesvg[width=0.8\linewidth]{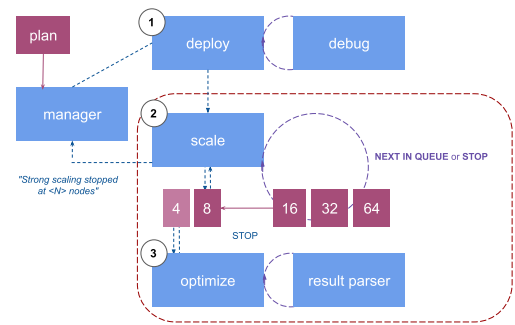}    
    \centering
    \vspace{1em}
  \caption{Scaling Study \normalfont Agentic Team for a hypothetical scaling study for four to 64 nodes (maroon boxes). The deploy agent \textbf{(1)} generates a manifest in a working state. The scaling agent \textbf{(2)} receives instruction to start the study, and updates the prompt to deploy at each requested size. At the smallest size, the scale agent directs execution to the optimization agent \textbf{(3)}, which decides on a configuration for subsequent sizes. At subsequent sizes, the scaling agent assesses the previous result, and decides to stop or proceed to the next size. Execution of subsequent sizes is done by the deploy agent. The scaling agent reports evidence and reasons for its decisions. Helper debug and result parsing agents assist primary step agents.}
  \label{fig:agentic-team}
\end{figure}

\smallskip
\noindent{\bf Result Assessment} 
\label{sec:result-assessment}
Across experiments, we are interested in the agents' choices for the Docker build, Kubernetes specifications, and the decision sequence to retry or stop. Each application will be run for 10 iterations, with a total number of retries of 10 for each step agent, and 15 retries for the manager. Each application will be allowed a maximum running time of 5 minutes, and a timeout will cancel the run and alert the agent that the application run has timed out.  Each result will be assessed by our team for quality, and agent logs will be inspected to learn how we can improve and tighten the orchestration. Our team has expertise in building and deploying \gls{hpc} applications in cloud environments \cite{sochat2025usabilityevaluationcloudhpc}. We evaluate:

\begin{itemize}
\item Overall performance of final optimized variant
\item Total time to complete each step
\item Number of attempts to successfully complete a step
\item Choices step agents take between and within steps
\item Dockerfile build logic correctness and completeness
\item MiniCluster design logic correctness and completeness
\item Decision of when to \emph{STOP} versus \emph{RETRY}
\item Reasons for failure
\end{itemize}

For failures, we will assess whether an additional prompt or requirement could better guide the agentic team to a successful outcome. We will run experiments in sessions, and make adjustments to prompts and software design to improve upon initial attempts.

\label{ref:token-counts}
\subsubsection{Token Counts and Completion Time}
A token is a chunk of text that an \gls{llm} processes that can be a single word, part of a word, punctuation, or a space. For Gemini, a token is about four characters \cite{gemini-tokens}. We are interested in the relationship between token counts for requests and responses, and time (seconds) for completion. It is not clear if providing more tokens in a request leads to longer processing time. It could be feasible to have a short response that requires more processing time by the \gls{llm}.

% \begin{comment}
% Note from V - other experiments of interest:

% \begin{itemize}
% \item allowing > 1 node (up to some max, maybe just 2). I'm interested in how it deals with scale but I don't want to spend too much.
% \item asking to minimize cost
% \item something that can derive a function we reuse?
% \item something with a job abstraction?
% \item giving more explicit instructions about the optimization logic
% \item testing the same app w/ and w/out exact parameters. Which case is more realistic? Which does the LLM handle better?
% \end{itemize}
% \end{comment}
\section{Results}
\label{sec:results}

We performed single node build, deploy, and optimization experiments for 4 proxy applications (AMG, LAMMPS, Kripke, Laghos) across 21 instance types on an autoscaling cluster followed by a multi-node (N=4) optimization study to add the networking benchmark OSU. We finished with a scaling study that extended the experiments to deploy and optimize across sizes. Best \gls{fom} results are shown in Table \ref{table:performance-results} and each experiment discussed below.

\label{sec:results-tokens}
\subsection{Tokens}

\begin{figure}[h!]
    \includesvg[width=\linewidth]{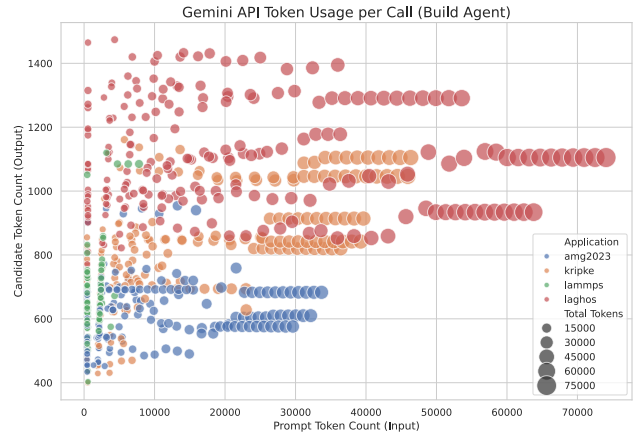}
  \caption{Gemini Token Counts for Build Agent. \normalfont The linear pattern of receiving back a similar or equivalent output token count reflects a likely convergence on response, either closing in on a solution or getting stuck in a broken state. The ease of generating a working build for LAMMPS (green) is reflected in requiring fewer total tokens.}
  \label{fig:gemini-token-counts-build}
\end{figure}

We analyzed the relationship between token counts and seconds to complete the response (Figure \ref{fig:gemini-token-counts-build}). We did not observe any strong relationship between prompt token count and candidate token count, however a pattern reflecting the difficulty of building the application was seen in our plot. Applications that are more difficult to build, either for a human or the \gls{llm}, produced a higher token count for prompts and candidates. The higher token count is caused by the library framework sending more error output, and receiving more text content. As an example, Laghos (green in Figure \ref{fig:gemini-token-counts-build}) was hardest to build, and has the highest prompt and candidate token counts. We also see ``stuck sequences'' in the data, or a sequence of prompts that are similar in count that likely received similar responses from the debugging agent. This pattern -- the inability of a request to return a working answer and trying something similar to return a similar result -- might be predictive of an oncoming failure state. It could also be indicative of fine-tuning a response that is closer to a successful solution. We hypothesize that the content of the prompts is similar due to what appears to be equivalent lengths. % More work is needed to better identify, label, and investigate this hypothesis.

\label{sec:results-single-node}
\subsection{Single Node Experiments}

\begin{figure}[b!]
    \centering
    \includesvg[width=1.0\linewidth]{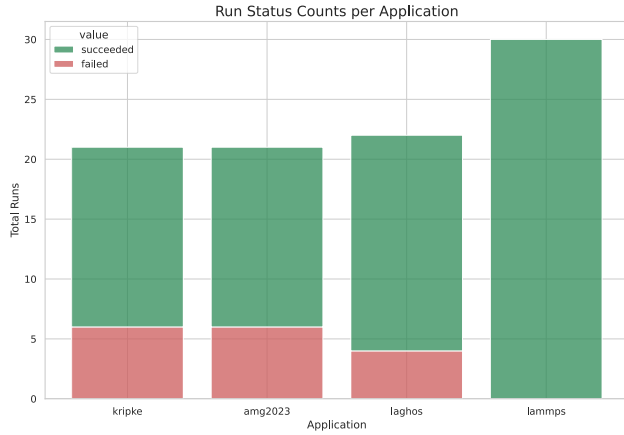}
  \caption{Agent Status. \normalfont A failed run indicates that the application did not make it to a successful completion (non-zero exit code). A successful run (green) indicates that a \gls{fom} was generated. LAMMPS had more runs due to initial testing.}
  \label{fig:application-status}
\end{figure}

We asked build, deploy, optimization, debugging, and parsing agents to work together with a manager to deploy AMG, LAMMPS, Kripke, and Laghos on one node. Overall success and failure rates are shown in Figure \ref{fig:application-status} and best performance metrics in Table \ref{table:performance-results}. A failure indicates that the pipeline was not completed through optimization because build or deploy did not complete successfully. Akin to the patterns described in Figure \ref{fig:gemini-token-counts-build}, we found that the difficulty an agent had with a build or deploy step matched our perceived human difficulty, with LAMMPS being the easiest to build and execute, and AMG, Kripke, and Laghos more difficult. Subjectively, more difficult builds require more dependencies, and the dependencies need customization with respect to build flags or settings. LAMMPS is easy to build with \emph{cmake}, while Laghos is much more challenging because it requires specific versions and flags for each of \emph{hypre}, \emph{mfem}, and \emph{metis}. % Results for different prompting strategies for one node, and for 4 nodes are described below.

% TODO for instance types figure we need a better illustration here - how to show breakdown by application but also in tems of agent type?

% separate multi node from single node, make own graphs. Separate into
% Be clear this is AS it is optimizing. 
% multiple strategoes for optimizing each application. Show how for each repetition of each strategy, 
% we want a lineplot of foms, where the axis is attempt, color of line is stragegy, we end at the max.

% rename multi node to llm-decision on multiple nodes
% choose the one with the maximum fom, and plot it change over time.

\begin{table*}[htbp]
  \centering
  \begin{threeparttable}
    \small
    \caption{Best Application Figure of Merits for Experiment Types}
    \label{table:performance-results}
    \begin{tabular}{llll}
      \toprule 
      Application & Experiment & {Best \gls{fom}} & Instance Type \\
      \midrule
      lammps  & Single Node (llm)  & $724.238$ katom steps/s & hpc7g.16xlarge \\
      lammps & Multi Node & $2.50$ Matom steps/s & m7g.16xlarge \\ 
      lammps & Scaling Study & $3.159$ Matom steps/s & hpc7g.16xlarge \\
      amg2013 & Multi Node & $6.132476\times 10^9$ nnz/s & m7g.16xlarge \\
      amg2013  & Single Node (llm) & $1.602391\times 10^9$ nnz/s & m7g.16xlarge \\
      amg2013 & Scaling Study & $6.424653\times 10^9$ nnz/s & hpc7g.16xlarge \\
      kripke & Single Node (llm) & $7.338117\times 10^{-10}$ (s/iter)/unknowns & c7g.16xlarge \\
      kripke & Multi Node & $7.795293\times 10^{-10}$ (s/iter)/unknowns & hpc7g.16xlarge  \\
      kripke & Scaling Study & $6.012693\times 10^{-10}$ (s/iter)/unknowns & hpc7g.16xlarge \\
      laghos & Single Node (llm) & $3.5295\times 10^2$ megadofs/second &  c7g.16xlarge \\  
      \bottomrule
    \end{tabular}
    \begin{tablenotes}[flushleft]
      \footnotesize % Use a smaller font for notes
      \item Best \gls{fom} reported by experiment type. For single-node experiments, the best strategy is also reported. For the scaling study, the best \gls{fom} across sizes is reported. \emph{llm}: llm decision, \emph{user}: user guided.
    Laghos units are in megadofs x time steps / second.
      % I've split your notes into \item for clarity, but you can also just have plain text.
    \end{tablenotes}
  \end{threeparttable}
\end{table*}
\vspace{1em}

% Note from V: for kripke multi node, it thought higher was better, so probably didn't optimize in the right direction. The fom is still alright but maybe not great.

%        app         experiment_type        best_fom
% 0   amg2023  user-provided-function         255.452
% 2   amg2023            llm-decision        3131.675
% 1   amg2023              multi-node        133008.7
% 3  amg2023    user-guided-function      15144090.0
% 5    kripke            llm-decision    0.0000000007
% 4    kripke              multi-node    0.0000000008
% 6    kripke    user-guided-function     0.000000021
% 12   laghos    user-guided-function  134.3420144394
% 11   laghos            llm-decision  352.9537111321
% 10   lammps    user-guided-function         355.524
% 9    lammps            llm-decision         724.238
% 7    lammps  user-provided-function          726.55
% 8    lammps              multi-node       2500000.0

% Note to Dan 10/24/2025: we have > 20 for lammps because I happened to run more iterations. If we want there to be exactly 20 we can randomly sample. Let me know if you want to do this.

\label{sec:results-instances}
\subsection{Instance Selection}

\begin{figure}[bh!]
  \centering
  \includesvg[width=1.0\linewidth]{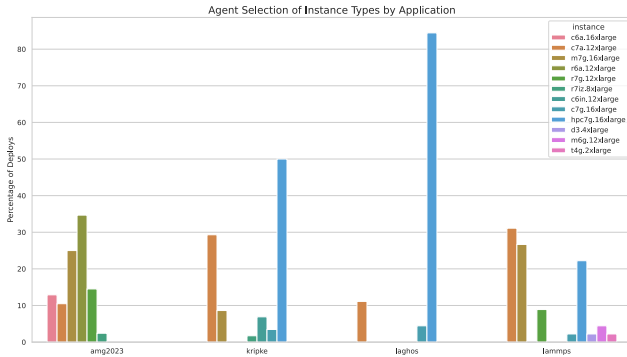}
  \caption{Agentic selection \normalfont across experiment runs for all applications. The agents heavily selected the \emph{hpc7g.12xlarge} for Laghos and Kripke, and did more sampling of the space for other applications.}
  \label{fig:instance-selection}
\end{figure}

Instance selection is the task for the \gls{llm} to choose a cloud instance type to assign a workload to. % which can come from the running cluster resources or an autoscaling cluster hint in the prompt details. 
% The deploy and optimize agents in our study could choose from a total of 21 instance types (Table \ref{table:instances}). 
For Laghos and Kripke, there was a preference for choosing the \emph{hpc7g.16xlarge}, with the instance type selected more than 50\% of the time (Figure \ref{fig:instance-selection}). Looking at the agent reported reasons, we see a pattern of the first optimization attempt ``selecting a 64-core, HPC-optimized instance.'' We hypothesize the choice is based on LLM training data that advertises the \emph{hp7g} as ideal for \gls{hpc} applications. When exploring multiple instance types, when the agent determines \gls{fom} has reached a limit for one instance type it often decides to test other types.

\label{sec:common}
\medskip
\noindent{\bf Common Errors} 
Agent responses that have common, general errors warrant an update to the prompt to avoid the erroneous states. An example is related to discovery of data and executables. If the build agent response did not place the final executable on the \emph{PATH} or add a comment to the Dockerfile about data file names and locations, it was unlikely for the deploy agent to be successful. Other consistent and surprising errors were not installing certificates and deleting binaries that were just built. Although these errors can be fixed by the debugging agent, it makes the loop tighter to prompt the agent to avoid them. 

\begin{figure}[hb!]
    \includesvg[width=1.0\linewidth]{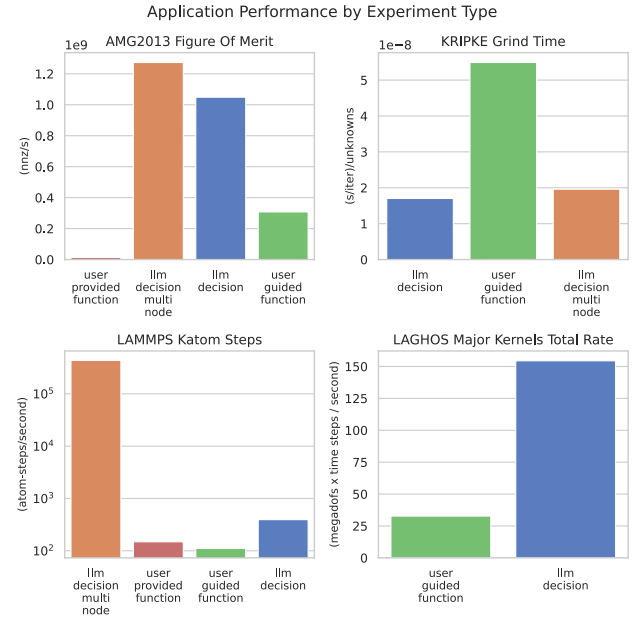}
    \centering
    \vspace{1em}
  \caption{Application best \gls{fom} by Experiment Type. \normalfont Allowing the \gls{llm} to decide how to optimize generally yielded the best results. For all \gls{fom}s, higher values are better, with the exception of Kripke (top right) where lower is better. The best strategy from single-node experiments was carried forward to multi-node experiments.}
  \label{fig:app-results}
\end{figure}

\label{sec:amg2023-result}
\medskip
\noindent{\bf AMG2013.}
Results for AMG are included in Figure \ref{fig:app-results}. Allowing the \gls{llm} to make a decision was the most successful strategy. The \gls{fom}s were $1.604893\times 10^9$, $1.417084\times 10^9$, and $8.048641\times 10^7$ for the \gls{llm} decision, user guided, and user provided functions, respectively. Despite requesting \emph{AMG2023} the agent always built the \emph{AMG2013} repository. This preference could derive from the \gls{llm}s training set. The difference is significant because the problem the agent was instructed to run (2) cannot be compared between variants.

We observed from agent output that the instruction from the user function was in direct conflict with what the \gls{llm} determined was best. We think this conflict results from the lack of fidelity of the hard-coded functions in modeling an environment. The user provided function did not work beyond the testing case because the optimization function produced a resource estimate that was too large, leading to execution that would run out of memory or time out. % We think that an expert on a specific instance type likely could write a well-performing function. 

We learned that without providing guidance for the problem size for the first deploy size, the agent always chose a problem size that was too large and would lead to the pod being \emph{OOMKilled}. Although the agent could adjust the problem size to be smaller each time, in practice it was not enough to get to a working result in under 10 attempts. We chose to advise the agent to choose a small problem size (no greater than 10) for testing of the initial deploy step. The subsequent information provided about how to optimize AMG for memory would then prevent the same error. If asking for a problem size that is too large is a common pattern, the agent might be given instructions for how to adjust problem sizes. Instead of a slow linear decrease, a binary search and then reaching a point when the \gls{fom} stops improving might be more efficient. 

Through a collaborative process of watching experiments run, we noticed that the key to AMG running well was to force OpenMP and any math libraries to use only one thread per MPI rank. While this insight could have come from an expert user, it was the \gls{llm} that reminded us of threading and we could reliably give the advice to the agent to make it the application run performant. 

% The second interesting thing is that at first it almost always wants to make a really big problem size. I tell it then what the CPU and memory are and that helps a bit, and it even adds comment to itself to reduce the value if the pod fails. It has a good sense of how to calculate the topology based on CPU, but isn't as good with guessing the problem size. But it's actually figuring it out - I'm watching it incrementally decrease the problem size with each failure, and that's a huge deal!
% I suspect this will be a common pattern in the future, and the issue with asking for more cpu than are available is that the job fails quickly enough that we can't easily grab the logs, and then the agent can't debug. I think I remember a command to get a log for a failed (deleted) pod - will look that up - we can use --previous.
% Amazingly, despite the above - it has gotten to running! It's smart enough to try a huge problem size and then reduce it after seeing an OOMKilled memory error. I'm impressed by that! It's gotten to the point in AMG where it will finish, but it could take a really long time. I want to add the ability to say "Please constrain your maximum running time to X" and then have it kill a working (running) job if that is violated to try again - it's the case where a "looks like could be successful, but may run until the end of time" should be considered an error. That should be easy to do - we can time the execution with the pull buffer, and generate an error state via our library that instructs the agent that the running time was too long.
\vspace{10pt}
\label{sec:lammps-result}
\noindent{\bf LAMMPS.} 
LAMMPS results are shown in the third panel of Figure \ref{fig:app-results}. While the user provided function mean value outperformed that of the user guided function, the variance across runs was twice as large, suggesting that the result is not consistent. The \gls{llm} decision outperformed both, leading to a \gls{fom} that was over 2x improved. The ease of building and deploying LAMMPS was reflected in our results. Builds did not require many debugging prompts (green in Figure \ref{fig:gemini-token-counts-build}) and the number of attempts was consistently fewer than five. The only application to consistently reach the maximum attempts was Laghos, which needed an increase to N=20. The ease of building mirrors the human experience. % For example, the \gls{llm} successfully completed a LAMMPS build always in under 5 attempts. 

% lammps   llm-decision                     724.238
%          multi-node                     2500000.0
%          user-guided-function             355.524
%          user-provided-function            726.55

\label{sec:laghos}
\medskip
\noindent{\bf Laghos.} 
Laghos was an exemplar of a build that required collaboration. In our testing runs, the agent was unable to build a container in the maximum attempts we allowed (N=10). Laghos is hard to build because several components (hypre, mfem, and metis) each require specific build flags and compile options. While the \gls{llm} might figure out a correct configuration eventually, the maximum attempts setting exists to disallow an unreasonably large number of builds. We fixed this by providing the agent with human expertise -- a description of the versions to build. With this expertise, the agent was successful every time. However, the components still took many attempts to get working together, reflected in a large number of prompts and responses for the build (Figure \ref{fig:gemini-token-counts-build}). We did not attempt a user provided function for Laghos due to the length of the build and previously poor performance with AMG.

\label{sec:kripke}
\medskip
\noindent{\bf Kripke.} 
The Kripke \gls{llm} decision outperformed the user guided function (Figure \ref{fig:app-results}), as a smaller grind time indicates higher performance. The \gls{fom}, Grind time per units of work (where smaller values are better) was order $10^{-8}$. 

% kripke   llm-decision               0.0000000007
%          multi-node                 0.0000000008
%          user-guided-function        0.000000021

% Note to Dan 10/24/2025: I was going to say amg was easy/hard to deploy, but I don't think it was easier or harder than any other, and I can't tell from looking at this result. I think (looking at the dots) this was a weird seaborn choice to make the line skinny at the top.

%\begin{figure}
%    \includesvg[width=\linewidth]{images/application-attempts.svg}
%  \caption{Agent Attempts. \normalfont We allowed for a maximum of 10 attempts for all applications with the exception of Laghos. Laghos was an immensely challenging build that we allowed for a maximum attempts of 20. LAMMPS was an easy image to build.}
%  \label{fig:application-attempts}
%\end{figure}

\label{sec:results-multi-node}
\subsection{Multi-Node Experiments}

A multi-node experiment requires extending the base container to include an OpenMPI build with libfabric enabled for \gls{efa}. We chose OpenMPI since it comes packaged with the \gls{efa} installer, and find that it works well across clouds and environments \cite{sochat2025usabilityevaluationcloudhpc}. We learned that asking for this additional dependency, especially for a multi-platform build, increases the time and difficulty of the build. Strategies for caching and reuse of images become increasingly important, and we adopted our strategy to ask the build agent to first build a base image with OpenMPI and libfabric, and then for the application builds we instructed it to use the shared base image.

\label{sec:amg2023-multi-node-result}
\medskip
\noindent{\bf AMG2013.} The optimized \gls{fom} for AMG on 4 nodes was $6.132476\times 10^9$ nnz/s on the \emph{m7g.16xlarge}. We observed that after the deploy agent deemed the application working in Kubernetes, the optimize agent increased the resources of the run to occupy the full node. While the agent was not instructed to, it chose to test two nodes before increasing to full size (N=4). We observed negligible improvement in performance, and attributed it to a small problem size. The next attempt scaled up to the full four nodes at maximum resources, and achieved a \gls{fom} of $6.13\times 10^9$. From this point, it increased problem size until performance degradation, which it attributed to ``exceeded the available memory or hit other scaling limits.'' % The agent strategy is rudimentary in that it did not test different processor topologies or environment variables for \gls{mpi}.

\label{sec:lammps-multi-node-result}
\medskip
\noindent{\bf LAMMPS.} LAMMPS optimized on 4 nodes resulted in a maximum \gls{fom} of 2.50 Matom steps/s. During optimization, the agent jumped immediately from one to the maximum of 4 nodes and 256 tasks, and increased problem size to \emph{-v x 32 -v y 32 -v z 16}. This initial attempt was too aggressive and timed out, leading the agent to adjust to a smaller problem size (\emph{-v x 24 -v y 12 -v z 12}) to establish a baseline (2.267 Matom/s). The final attempt increased the problem size to the deemed optimized \gls{fom} (\emph{-v x 30 -v y 15 -v z 15}).

\begin{figure}[b!]
    \includesvg[width=\linewidth]{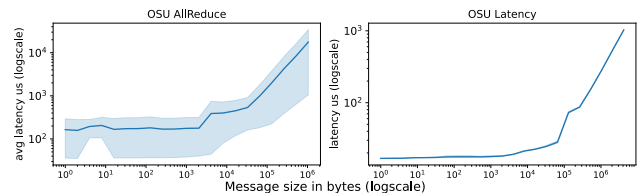}
  \caption{OSU Benchmarks \normalfont AllReduce (left) and Latency (right).}
  \label{fig:osu}
\end{figure}

\label{sec:osu-result}
\medskip
\noindent{\bf OSU Benchmarks.} OSU Benchmarks executed successfully without prompt tuning. OSU Latency had performance comparable to previous performance studies using \gls{efa} \cite{sochat2025usabilityevaluationcloudhpc} and OSU AllReduce showed high variability across message sizes (Figure \ref{fig:osu}). Agents successfully enabled collective and point to point communication.

\label{sec:kripke-multi-node-result}
\medskip
\noindent{\bf Kripke.} Kripke Grind time optimized to run across 4 nodes was comparable to one node, with a slightly higher time to perform a unit of work ($8\times 10^{-10}$ and $7\times 10^{-10}$ for each of 4 and 1 nodes, respectively). The decrease in performance likely results from the added need for network communication. Since it could be challenging in our experience \cite{sochat2025usabilityevaluationcloudhpc} to get Kripke parameters correct, we requested the \gls{llm} to test running \emph{kripke} without parameters.

% Kripke, we had an inflection point discovery in which consistent failures started to succeed when we requested the \gls{llm} test on a single node without parameters. Attempting to test on one node with parameters would result in an invalid topology. Once the application executed and the scale-up completed, the \gls{llm} added correct topology parameters.

% Note from V: why might that be?

% kripke   llm-decision               0.0000000007
%          multi-node                 0.0000000008
%          user-guided-function        0.000000021

\label{sec:results-scaling-study}
\subsection{Scaling Study}

\begin{figure}
    \includesvg[width=\linewidth]{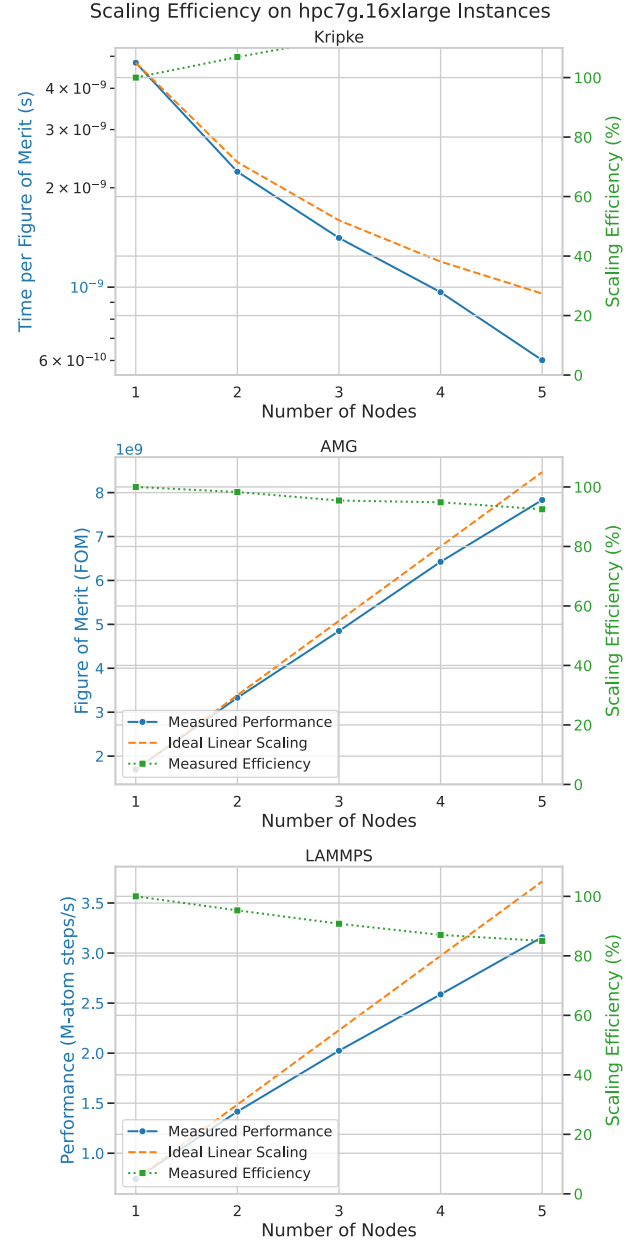}
  \caption{Application Scaling Efficiency \normalfont \emph{hpc7g.16xlarge} Instances for each of Kripke, AMG, and LAMMPS. An initial successful deployment using a small problem size is done first, then an optimization on one node, and pinning of the decided problem size up to the maximum size. For all applications, higher is better, with the exception of Kripke, where lower is better. The decision to stop/proceed is done by an agent.}
  \label{fig:scaling-study}
\end{figure}

We performed an agentic scaling study for LAMMPS, Kripke, and AMG. We had wanted to use a large cluster with 32 nodes, but were only able to provision 5 \emph{hpc7g.16xlarge} nodes over the course of a week. The \gls{fom}s are shown in Figure \ref{fig:scaling-study}.

We closely monitored each study for correctness. We learned that agents need specific instructions about rules for each step. For example, the optimization agent must be given instructions to keep trying until a condition like a timeout, and whether to minimize or maximize \gls{fom}. The specific size and history of \gls{fom} progression at each size is helpful. In early testing, we observed that when we got application-specific prompts correct, the study ran smoothly, meaning observing subsequent attempts were carefully changing the application in a way we thought reasonable without making mistakes (e.g., stopping too early, changing a problem size incorrectly).

We carefully monitored each decision and step in the scaling experiments, and learned how to handle unexpected edge cases and how to improve usability. For example, we had a case where the job controller did not recreate a Pod that did not bootstrap correctly with Flux. The ideal outcome would be to delete the Job, disregard the run, and try again. We added support for this case, a ``Lost'' status that can be triggered by unexpected deletion of the MiniCluster. We also learned that it was strategic to prompt the deploy agent to give unique names to Job with an incrementing number to avoid naming conflicts and to provide a live record via completed Pods for a human participant to inspect. Finally, we identified a tradeoff between choosing fewer edges in the graph and adding edges for error debugging. During the single-node optimization step, it would require fewer edges to give the optimization agent information about a failed run to debug. However, this practice violated our principle to have each agent perform one function, and in practice we saw better debugging by exiting the scaling and optimization loops and sending the diagnostic logs to the debugging agent to synthesize the problem and redeploy back into a working state.

\label{sec:lammps-scaling-result}
\medskip
\noindent{\bf LAMMPS.}  The LAMMPS optimization agent running on five nodes reached a maximum \gls{fom} of 3.159 Matom steps/s. We observed the scaling agent carefully increasing the problem size until the \gls{fom} stopped improving, even reporting on the percentage improvement between attempts at times. We observed that agent responses included correctly calculated percent changes in \gls{fom} output, suggesting that Gemini uses supporting tools on its backend. The agent step stopped and returned the result when the execution timed out. Attempts for 2,3,4, and 5 nodes followed instructions to use the fixed problem size \emph{18 x 18 x 18}. Comparing the single node optimization to the single node study, the \gls{fom} was slightly improved, hitting 743.061 katom-step/s, an improvement from 724.238 katom/steps using a problem size of \emph{-v x 16 -v y 16 -v z 16}. 

\label{sec:amg-multi-node-result}
\medskip
\noindent{\bf AMG2013.}  The scaling study produced a best \gls{fom} of $6.424653\times 10^9$ nnz/s for AMG on 5 nodes. For optimization on one node, the agent immediately increased the problem size to \emph{-n 270 270 270}, which resulted in the Pods being \emph{OOMKilled}. Subsequent attempts slowly decreased the size \emph{ (270, 215, 150, 100)} until a successful run was found \emph{ (-n 80 80 80)}. The agent then reversed the pattern, increasing the problem size slowly until the ``optimal'' \emph{ (-n 90 90 90)}, where the agent stopped due to OOM. While other applications typically had one run for sizes 2-5, the agent acted differently here, trying several variants of the processor topology for several sizes.

% I'm not sure if the above is too human sounding. I'm getting tired and having less of a hard boundary to decide. My agentic brain is failing me!! OMG I'm an agent too. beep boop.

% Question for Dan - where / how do we want to discuss the scaling agent decisions? E.g., each node count has something like:

% According to the principles of strong scaling (Amdahl's Law), for a fixed total problem size, the time to solution should decrease as the number of processing elements increases, up to a point where communication and other overheads begin to dominate. Our Figure of Merit (FOM) appears to be a measure of time per unit of work, so we expect it to decrease with added nodes. This initial run establishes the baseline performance. As noted in performance studies like those detailed in the Kripke user manual and related publications from Lawrence Livermore National Laboratory, observing the inflection point where performance gains diminish is the primary goal of such a study. We must proceed to the next size to begin observing this curve.

\label{sec:kripke-multi-node-result}
\medskip
\noindent{\bf Kripke.} Kripke's best \gls{fom} for the scaling study at 5 nodes was a Grind time of $6.012693\times 10^{-10}$ (s/iter)/unknowns. The optimization agent started at a small number of cubic zones (with edge length 32, 48, 64, 80, 96, 112, 120, and 128), and methodically increased until the execution ran out of memory. Size 128 failed, and size 120 was determined optimal. At sizes 2, 3, 4, and 5 nodes, the same problem size was returned in the response to run, and responses tested three different configurations of processor topology to maximize the \gls{fom}. We observed a linear speedup that would be possible if the problem size per core became small enough to fit into the CPU's fast cache (96 MB L3 cache), reducing time waiting for data from the slower main memory. As an example, our problem running on 5 nodes of the \emph{hpc7g.16xlarge} had a fixed total problem size of 103,555 MB. As we scaled up to 5 nodes, the data footprint per core decreased from 1,618 MB to 324 MB. While the full domain per core still exceeds the cache capacity, the Kripke wavefront of actively computed data is much smaller. By scaling to five nodes, this active wavefront is small enough to be cache-resident. We think that the transition to using the cache shifted the performance bottleneck from slow main memory to the faster L3 cache, resulting in an efficiency greater than 100\%. We view this optimization and scaling effort as highly successful, as tuning happened for problem size and processor topology.

% Then need to look at assessment points below (write up section)
% also need to add in scaling agent logic - is it right?

\label{esult:result-assessment}
\subsection{Result Assessment}

For our result assessment, we went through each study reported in Table \ref{table:performance-results} and qualified good and bad design choices. We were interested in Dockerfile and MiniCluster configuration logic correctness and design, and parameter or other environment choices. We include a summary of common good and bad choices for applications in Table \ref{tab:features-summary}.

% Note to Dan 10/24/2025: I couldn't figure out how to fix the cropping of Good / Bad. (Category column) so I just deleted the column names and the categories, and the green/red implies good/bad.
\begin{table}[ht]
\centering
\small % Slightly smaller font
\setlength{\tabcolsep}{4pt} % Reduce padding between columns
 \caption{Evaluation of Practices in Experiments}
  \begin{threeparttable}
    \begin{tabular}{lcccc} 
    \toprule
    \textbf{Choice} & \textbf{LAMMPS} & \textbf{Kripke} & \textbf{Laghos} & \textbf{AMG} \\ % Shortened AMG2013 to AMG
    \midrule
    \rowcolor{green!15}
    Non-interactive setup & 2 & 2 & 1 & 2 \\
    \rowcolor{green!15}
    \texttt{apt clean} usage & 2 & 1 & 1 & 0 \\
    \rowcolor{green!15}
    Allow root / envars & 1 & 1 & 1 & 1 \\ % Shortened description
    \rowcolor{green!15}
    Shallow Git clone & 2 & 1 & 0 & 0 \\
    \midrule
    \rowcolor{red!15}
    Monolithic layers & 2 & 1 & 1 & 2 \\ % Shortened description
    \rowcolor{red!15}
    Network assumptions & 1 & 0 & 1 & 1 \\ % Shortened description
    \bottomrule
    \end{tabular}
    \begin{tablenotes}[flushleft]
        \footnotesize
        \item[*] Good practices (green) vs. bad (red). 0-2 scale.
    \end{tablenotes}
  \end{threeparttable}
  \label{tab:features-summary}
\end{table}

\label{sec:app-specific-assessment}
\medskip
\noindent{\bf Application-specific Choices.} 
Despite not being advised about networking, the LAMMPS build agent set environment variables to assume ethernet, which we considered a non-ideal choice in the case that the user deployed a different device. % The multi-node build had similar patterns of build logic, and we generally had no negative criticism for the MiniCluster configuration files.
Most builds, with the exception of Kripke, had too many distinct layers that should have been bundled into one layer.
% The OSU Benchmarks require careful choices with respect to number of tasks (e.g., point-to-point tests must be run on 2 nodes with exactly two processes) and the agents consistently produced the correct MiniCluster. Akin to LAMMPS, the Dockerfile had layers that included too many logical units, but best practices to run in non-interactive mode were used. The agents consistently chose Ubuntu 22.04. We think the choices of container bases reflect the \gls{llm}s training set.
% Unlike OSU and LAMMPS, the build of Kripke was done as an isolated logical unit to clone, compile, and cleanup, which we consider best practice. 
% Environment variables to allow OpenMPI to run as root were seen, and considered good practice provision root as the default user. Network parameters were included as advice to the user as comments.
% The Laghos build was the most challenging of the set, requiring exact versions in the prompt. 
The Laghos build would not have been possible without very strict prompt input.
% AMG2023 requested builds consistently built AMG2013, which was surprising due to being older, and the license less permissive. The agent responses included good choices to add environment variables for running as root. 

\section{Discussion}
\label{sec:discussion}

In this work, we demonstrate the ability of an agentic team to semi-autonomously execute the entire workflow to build, deploy, and optimize \gls{hpc} applications. There are several topics for discussion.

\smallskip
\noindent{\bf \gls{llm} Collaboration} 
Collaboration between \gls{llm} agents and humans is important. We learned about a new CPU request unit, a millicore \cite{millicore}. From observing experiments, we learned that \gls{llm} agents require succinct prompting, and user validation along the way ensures the agent is parsing correctly. % The agents would not have reached successful outcomes without our insights, and often our insights were not possible without observing the \gls{llm} behavior.
% We discovered there are many cases when the \gls{llm} makes an error that could easily be avoided with expert guidance, and the collaboration extends beyond prompting to interacting with the \gls{llm} during execution. While the \gls{llm} is capable of acting autonomously, a human can validate a pipeline along the way. In the case of generating a method to derive the result that can be cached or saved for subsequent operations (e.g., a regular expression) the need for human support is less frequent. 
Asking the \gls{llm} to provide rationale for choices gives insight to see flaws in logic. % or missing information in the logic of the response. This transparency guides prompt adjustment.

\smallskip
\noindent{\bf Self-healing Systems} 
The choice of number of retries to allow an agent, and the strictness of guardrails provided must strike a balance that allows the agent to explore, but up to reasonable constraints. In any case, the entire system must be able to self-heal, or come back from erroneous state or respond to context drift \cite{context-drift}. While there are many strategies for self-healing systems \cite{nutalapati2022self}, we chose a simple strategy to ensure several means to restart: a debugging agent can decide at any time to return to the manager, or the same outcome can result if the number of maximum attempts is exceeded. We also give the agent the option to return to a human. % where a summary is presented to the human experimenter, and they can respond to the \gls{llm} with textual advice. 

% Another feature of a self-healing system is an ability to respond to agentic context drift \cite{context-drift} into more erroneous state. We noticed, for example, that the result parser might produce a regular expression that was close to correct, but did not match the log. Receiving feedback about the incorrect generation would then push the regular expression in an entirely different (and often wrong) direction. We decided to attenuate the behavior by allowing a human user to intervene with an instruction at the 5\textsuperscript{th} attempt, and resetting history at the the 10\textsuperscript{th} attempt. In the case of the example, the user could provide details that the first regular expression tried was the correct item to parse and very close, and to focus on it.

\smallskip
\noindent{\bf Temporal Pinning and Dependency Versions} 
Care must be taken when advising the agents to use specific versions of software or dependencies. We discovered that asking for a specific version of a library tended to pull versions of other software and base images toward specific points in time (e.g., asking for OpenMPI 4.1.2 produced builds with Ubuntu 20.04 and libfabric 1.13 that is no longer compatible for \gls{efa}). % While some cases of this behavior could be acceptable, in this case, the old libfabric did not function with the newer hardware on the instance. When not prompted to use a specific version, we tended to see more up to date versions (Ubuntu 22.04 and OpenMPI 5.x) that likely reflected a more recent bulk of textual data from training. 

\smallskip
\noindent{\bf Decision to STOP} 
Asking the \gls{llm} the algorithm it was using to determine when the application was optimized gave us insight into observed behavior. The \gls{llm} often reported a greedy hill-climbing approach, where it would evaluate the last choice, determine if the change improved or was detrimental to performance, and then would either step back to a previous configuration or continue in the same direction. % The \gls{llm} seems to move toward a local optimum and stop when there is not obvious improvement and other resource tweaks to try. While the \gls{llm} would often choose a different instance type to start and change once or twice, it was not common for the \gls{llm} to want to try multiple instance types. We observed that the \gls{llm} agent always attempted to produce reasonable logic (comments in the configuration files) for making a choice, even if the logic was wrong. 
Many of the stopping conditions we saw as premature, and we think more human guidance about an approach would be needed. % For example, we might instruct the agent to make a particular number of configuration changes before deciding to stop, or reaching a performance plateau.

% TODO to include
% number of times for each it changed instance types
% total number of times it tried before determining it was optmized.

\smallskip
\noindent{\bf Optimization Goals} 
Our most successful experiments were specific about optimization. Minimizing walltime without specifying the problem size is insufficient because the \gls{llm} can simply run a small problem more quickly. Not providing a small test case can allow the \gls{llm} trying to run problems that are too large too quickly. Not providing a strategy or function for incrementing resources can lead to jumps that are too large or too small for the number of attempts allowed. Best practices for individual steps can often make the step more challenging for the agent. For example, while it is the case that a multi-stage build is a best practice for Docker images to reduce image size, in practice it made it more challenging for the build agent, because library dependencies or required data were less likely to be carried forward between stages. % Allowing for the build step to build on two platforms, on the other hand, increased build time without adding additional error. We chose to build on two platforms to double the number of possible instances the deploy step could choose from.

%\smallskip
%\noindent{\bf User-provided Functions} 
%Our experiments demonstrated that the \gls{llm} agents outperformed our functions, likely due to not having rigidity and our lack of expertise of applications running on specific instance types. We observed that while the user provided function did not yield results for AMG, it outperformed the user guided function for LAMMPS. We think that a scoped, user provided function could be comparable to an \gls{llm} agent decision if it is written by an expert for the environment. This is work we anticipate doing.

\smallskip
\noindent{\bf Emphasis in Prompts} 
We learned early on that without emphases added to our prompts, (e.g., \emph{You MUST}) the \gls{llm} would often not follow the instruction. Adding capital letters made a large difference when we absolutely did not want a behavior. When the \gls{llm} made a decision that went against one of our requests, it added code commentary that would justify the choice.

% 10/31/2025 Dan - This is redundant (we mentioned this previously) not sure if we should just remove this section - if you agree PLEASE USE YOUR GIGANTIC DELETE BUTTON!

\smallskip
\noindent{\bf Experiential Learning} 
The most dangerous failure cases were those that the agent could not learn from (e.g., a timeout where the agent does not get feedback) leading to guessing behavior. % Our follow up work provides tools for agents to better inspect the environment. % It would need to infer which of its choices led to the slow application execution, and try something differently. Only by trial and error and observation that threading made a huge difference in avoiding timeouts and increasing problem size that we learned to explicitly provide this advice to the \gls{llm}. Another fruitful strategy was the option to return to the manager. In this case, the manager was required to summarize the issue to return to an agent, and we found consistent issues in the summary messages that we were able to integrate into prompts. After this change, there were no subsequent returns to the manager due to the success of the deployment.

\subsection{Future Work and Limitations}
We have completed follow up work that extends this framework to include a hierarchical agentic server, and orchestration with \gls{mcp}.

We recognize that other LLM \gls{api} endpoints could have been used given equivalent opportunity. We recognize that more research is needed to better understand when and how a human should intervene.

\section{Conclusion}

Agentic frameworks are the future for not just the \gls{hpc} community, but the entire software ecosystem. Autonomous, converged \gls{hpc} infrastructure that can collaborate with humans to orchestrate the entire lifecycle of scientific workloads is becoming feasible.  From autonomous task execution to failure recovery and \gls{ai}-supported scheduling techniques, these powerful transitions will fundamentally change our daily work lives and how scientists, engineers, and developers pursue scientific discovery. We look forward to a new era of faster, more accurate, and more accessible computing for the next generation of HPC.

\begin{acks}
% Thank you to my mujuh and thank you to my fajah for deez arms, legs, and sentient meat that sits between my ears that is pretty good but would taste absolutely terrible on a burger. Thank you for these last years that have felt easy and full of strength and joy.
This work was performed under the auspices of the U.S. Department of Energy by Lawrence Livermore National Laboratory under Contract DE-AC52-07NA27344 (LLNL-TR-2013651). % and was supported by the LLNL-LDRD Program under Projects No. 22-ERD-041 and 24-SI-005 (LLNL-CONF-XXXXXX).
\end{acks}

\bibliographystyle{ACM-Reference-Format}
\bibliography{references}

% \appendix
% \input{sections/appendix}

\end{document}